\documentclass[ preprint, superscriptaddress,  amsmath,amssymb,  aps,  prb
]{revtex4-1}%
\usepackage{graphicx}
\usepackage{bm}
\usepackage{hyperref}
\usepackage{amsmath}
\usepackage{amsfonts}
\usepackage{amssymb}%

\begin{document}
\title{Low-threshold directional plasmon lasing assisted by spatially coherent
surface plasmon polaritons}
\author{Yang Cao}
\affiliation{Department of Physics,Tongji University,Shanghai,200092,China}
\affiliation{Key Laboratory of Advanced Micro-structure Materials, MOE, Department of Physics, Tongji University, Shanghai 200092, China}
\author{Zeyong Wei}
\affiliation{Department of Physics,Tongji University,Shanghai,200092,China}
\affiliation{Key Laboratory of Advanced Micro-structure Materials, MOE, Department of Physics, Tongji University, Shanghai 200092, China}
\author{Wei Li}
\affiliation{State Key Laboratory of Functional Materials for Informatics,Shanghai
Institute of Microsystem and Information Technology,Chinese Academy of Sciences,200050,Shanghai,China}
\author{Anan Fang}
\affiliation{Department of Physics,Hong Kong University of Science and Technology,Clear
Water Bay,Kowloon,Hong Kong,China}
\author{Hongqiang Li}
\email{hqlee@tongji.edu.cn}
\affiliation{Department of Physics,Tongji University,Shanghai,200092,China}
\affiliation{Key Laboratory of Advanced Micro-structure Materials, MOE, Department of Physics, Tongji University, Shanghai 200092, China}
\author{Xunya Jiang}
\affiliation{State Key Laboratory of Functional Materials for Informatics,Shanghai
Institute of Microsystem and Information Technology,Chinese Academy of Sciences,200050,Shanghai,China}
\author{Hong Chen}
\affiliation{Department of Physics,Tongji University,Shanghai,200092,China}
\affiliation{Key Laboratory of Advanced Micro-structure Materials, MOE, Department of Physics, Tongji University, Shanghai 200092, China}
\author{C.T. Chan}
\affiliation{Department of Physics,Hong Kong University of Science and Technology,Clear
Water Bay,Kowloon,Hong Kong,China}

\pacs{42.55.-f,42.25.-p,73.20.Mf,78.20.Ci}

\begin{abstract}
We theoretically propose directional, low-threshold plasmon lasing in both the
near-infrared and visible wavelengths by utilizing spatially coherent surface
plasmon polaritons on a meta-surface. The gain strength required for threshold
lasing can be tuned down to a large extent through compatible structural
parameters. Our calculations show that no more than 65 cm$^{-1}$ at 193.5 THz
(1.55 $\mu$m) or 267 cm$^{-1 }$at 474THz (0.633 $\mu$m) of gain coefficient is
sufficient to compensate for the dissipation of metal films for threshold
lasing; these values are smaller than any reported studies at the same
frequencies. These findings present a planar solid-state route for plasmon
lasing that is highly efficient and spatially coherent.

\end{abstract}
\maketitle

\section{Introduction}

In recent years, nanophotonics has attracted considerable attention as one of
the emerging fields of modern science and technology. Active devices,
especially coherent sources with high efficiency, are of great importance for
the development of nanophotonics. The amplification of the evanescent field
and strong confinement of light in a subwavelength scale are crucial for
active nanophotonic devices to achieve superior performance. Surface plasmon
polaritons (SPPs) provide a feasible route for manipulating light beyond the
diffraction limit. In 2003, Bergman and Stockman proposed surface plasmon
amplification by stimulated emission of radiation (SPASER)\cite{1,2}. A
spaser---referring to the accumulation of a large number of SPPs in a single
mode assisted by a gain medium---is precisely the plasmonic counterpart of the
nano-laser. In principle, one of the most important applications of SPP
amplification is plasmon lasing\cite{3,4}. As energy dissipation in metals
significantly reduces the lifetime of SPPs in the infrared and visible
frequencies, the gain of the active medium must prevail over the dissipation
of metals in order to sustain the survival of a spaser. Many efforts related
to SPP amplification have been devoted to loss compensation in
metal-dielectric structures\cite{5,6,7,8}. The first experimental verification
of the spaser used gold nano-particles encapsulated in dye-doped silica
shells, which support a collective SPP mode with a high quality factor
(Q-factor)\cite{9}.

Planar design is preferred for the realization of spasing as well as plasmon
lasing for the sake of fabrication convenience and device integration.
However, as a finite-sized planar structure is usually incapable of strong
light confinement, a high-gain medium is required in order to compensate for
the radiation and damping losses from the planar structure. One solution to
this difficulty is to utilize \textquotedblleft trapped mode\textquotedblright%
\ of a plasmonic meta-surface\cite{10}. A split ring resonator with weak
asymmetry (ASR) possesses a leaky mode that is poorly coupled with free-space
photons. A plasmonic metamaterial slab composed of an array of such ASRs will
facilitate SPP amplification and lasing with low-gain material when the leaky
mode is excited. Subsequent experimental investigation showed that the loss
from the nano-sized metallic ASRs could be compensated for by the gain of
quantum dots\cite{11}.

It is worth noting that a magnetic meta-surface can support spatially coherent
SPPs which are of high Q-factor as well. The subtle concept, in contrast to
the \textquotedblleft trapped mode\textquotedblright\ which is derived from
dark mode of a single resonator, comes from the SPPs in harmonic
mode\cite{12}. These spatially coherent SPPs are actually the leaky waves of a
slab waveguide which can not be determined by a single resonator. They must be
in harmonic mode to sustain a leaky wave character, giving rise to a high
Q-factor which can be of one or two orders larger as compared to that of the
SPPs in fundamental mode. In this paper, we theoretically propose that such
spatially coherent SPPs present a practical solution for low-threshold plasmon
lasing in both the near-infrared and visible regimes. The gain coefficient
$\alpha$ of the active medium for threshold lasing can be tuned down by
structural parameters to a large extent, even down to values much smaller than
those of semiconductor quantum wells. Two model samples are presented with
structural parameters that are feasible for practical fabrication. The
threshold gain coefficient $\alpha_{\text{th}}=$65 cm$^{-1}$ at 193.5 THz
(1.55$\mu$m) is much smaller than those of other lasing systems reported in
the same infrared frequency region\cite{10,13,14}. The low threshold arises
from the high Q-factor of the SPPs in harmonic mode and from the strong
confinement of the local field in the active layer, both of which are not easy
to establish in a planar solid-state system. The high directionality of the
lasing beam with the full width at half maximum (FWHM) of $\Delta
\theta=4^{\text{o}}$ in E-plane ($\varphi=0^{\text{o}}$) and $\Delta
\theta=16^{\text{o}}$ in H-plane ($\varphi=90^{\text{o}}$) is verified by
finite-difference-in-time-domain (FDTD) simulations, where $\theta$ and
$\varphi$ are the polar angle and azimuthal angle.

\section{Reflectance amplification assisted by spatially coherent SPPs}

Our planar lasing system is schematically illustrated in the inset of Fig.
1(a). Lying on the $xy$ plane, the slab comprises an upper layer of a metallic
lamellar grating with a thickness of $t$, a metallic ground plane, and an
active layer with a thickness of $h$ sandwiched between them. The metallic
strips with a width of $a$ are separated by a small air gap $g$, giving rise
to a period of $p=a+g$ for the lamellar grating. The geometric parameters of
our model in the infrared region are $t=0.05\mu$m, $h=0.2\mu$m, $a=1.1\mu$m,
$g=0.05\mu$m, and $p=a+g=1.15\mu$m.

The SPP resonances on the slab can be characterized by the amplification of
reflection coefficients. The metals in our system are assumed to be silver
with permittivity defined by a Drude model$\cite{15}$, $\varepsilon_{m}%
(\omega)=\varepsilon_{\infty}-\omega_{p}^{2}/(\omega^{2}+i\omega\gamma)$ where
$\varepsilon_{\infty}=3.7,$ $\omega_{p}=13673$THz, $\gamma=27.35$THz. And the
dielectric layer is assumed to be active with a complex permittivity
$\tilde{\varepsilon}={\varepsilon}^{\prime}+i{\varepsilon}^{\prime\prime}$, in
which the imaginary part ${\varepsilon}^{\prime\prime}$ is positive. We
perform a rigorous coupled-wave analysis$\cite{16,17}$ to calculate the
coefficient $r_{m}(\theta,f)$ of the $m^{\text{th}}$ order of the reflected
Bloch wave components, in which $\theta$ and $f$ denote the incident angle and
the frequency of plane wave incidence. The incident plane wave is transverse
magnetic (TM) polarized with the magnetic field parallel to the $y$ direction
and in-plane wave vector $\vec{k}_{\parallel}=k_{0}\sin\theta\hat{e}_{x}$,
where $k_{0}$ is the wave vector in a vacuum. The gain strength of the active
layer is defined by a coefficient, $\alpha=(2\pi/\lambda)Im(\sqrt
{{\varepsilon}^{\prime}+i{\varepsilon}^{\prime\prime}})$. In the calculations,
different values of gain strength are taken into account by varying the
imaginary part ${\varepsilon}^{\prime\prime}$ of the complex permittivity
while keeping the real part fixed at ${\varepsilon}^{\prime}=1.7$.

\begin{figure}[pb]
\begin{center}
\includegraphics[
width=8cm
]{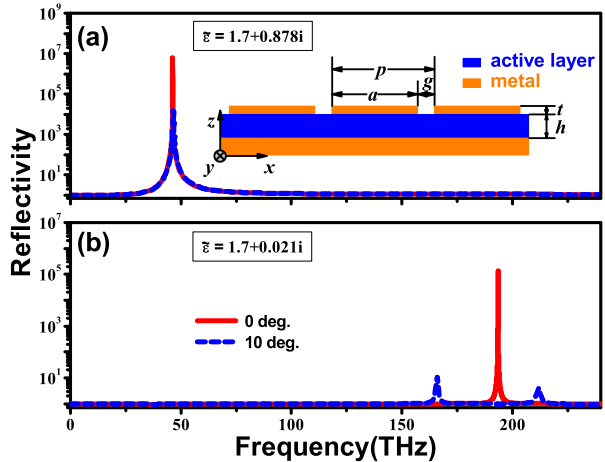}
\end{center}
\caption{Reflection spectra under TM polarized plane wave incidence at =
$0^{\text{o}}$ (red solid lines) and $10^{\text{o}}$ (blue dashed lines) with
different values for permittivity of the active medium: (a) $\tilde
{\varepsilon}={1.7}+0.878i$, (b) $\tilde{\varepsilon}={1.7}+0.021i$. The inset
in Fig. 1(a) shows the schematic geometry of our sample model.}%
\end{figure}

Figure 1 shows the calculated reflection spectra in the 0$^{\text{th}}$ order
under an incident angle of $\theta=0^{\text{o}},$ $10^{\text{o}}$ and an
imaginary part of the permittivity of the active layer of ${\varepsilon
}^{\prime\prime}=0.878,$ $0.021$. When the gain strength is strong
(${\varepsilon}^{\prime\prime}=0.878$), there exists an amplification peak
fixed at 46.3 THz ($\lambda=6.48\mu$m) that is almost independent of the
incident angle. In contrast, when the gain strength is weak (${\varepsilon
}^{\prime\prime}=0.021$), the amplification peak at 46.3 THz disappears, and a
new peak emerges at 193.5 THz ($\lambda=1.55\mu$m) under normal incidence.
Under oblique incidence, the emerging peak splits into two peaks with varying
frequencies that are sensitive to the incident angle; see, for example, the
dotted line in Fig. 1(b) for the two peaks at 166.1 THz and 211.9 THz with
$\theta=10^{\text{o}}$. The angle-dependent amplification peaks are obviously
different from the angle-independent peak in regards to the physics of their
origin. Their time-reversed counterparts are the angle-dependent and
angle-independent absorption peaks that we have investigated before$\cite{12}%
$. An angle-independent peak, requiring a sufficiently large gain/absorption
coefficient of the dielectric layer, comes from the excitation of a localized
SPP resonance in a fundamental mode that is dominated by the 0$^{\text{th}}$
order guided Bloch mode in the dielectric. The angle-dependent feature comes
from the spatially coherent SPP resonance in a harmonic mode that is dictated
primarily by the $\pm$1$^{\text{st}}$ orders of the guided Bloch mode in the dielectric.

\begin{figure}[ptb]
\begin{center}
\includegraphics[
width=8cm
]{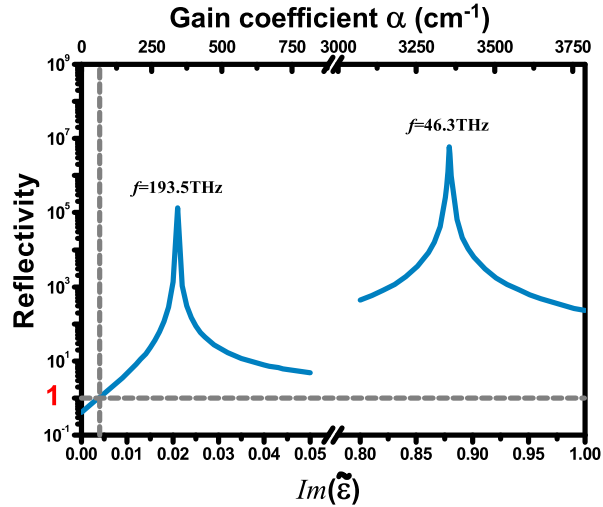}
\end{center}
\caption{Reflection amplification at 193.5 THz and 46.3 THz as a function of
gain strength of the active layer.}%
\end{figure}

The reflection amplification of the two resonant peaks as a function of the
gain strength in the active layer is plotted in Fig. 2. It should be noted
that a rather low gain strength $\alpha_{\text{th}}=65$cm$^{-1}$is sufficient
to compensate for the loss of metals and to sustain the amplification of the
guided waves in the active layer as well as the reflectance amplification at
193.5 THz. The amplified reflectance increases exponentially with the increase
of $\alpha$, until a maximal amplification of about $1.4\times10^{5}$ is
reached at an optimal value of $\alpha_{\text{opt}}=342$cm$^{-1}$. In
contrast, we see from Fig. 2 that the reflectance amplification at 46.3 THz
requires much stronger gain strength, which is unattainable for most of the
existing active media at this frequency.

Additional calculations indicate that the gain strength required for the
reflection amplification at 193.5 THz can be reduced further, provided that a
smaller air gap or a thinner dielectric layer which matches the gain strength
is adopted. The property provides us with a good opportunity for finding a
balance between ease of fabrication and attainable gain strength of the active
layer. Usually thermal conductivity of active medium becomes larger with
respect to higher temperature. A weaker gain strength required for threshold
lasing also implies that we can find an appropriate structure for plasmon
lasing that is robust against thermal noise, or intuitively that the thermal
noise is somewhat \textquotedblleft suppressed\textquotedblright\ by spatially
coherent SPPs. This is advantageous for room-temperature plasmon lasing as one
goal being pursued. At this stage, it is imperative for us to apply the
spatially coherent SPP modes for low-threshold plasmon lasing as such a SPP
mode with a long lifetime, which is compatible to the structural parameters,
will create extensive adaptability in material selection and feasible way for
practical implementation of plasmon lasing.

\begin{figure}[ptb]
\begin{center}
\includegraphics[
width=8cm
]{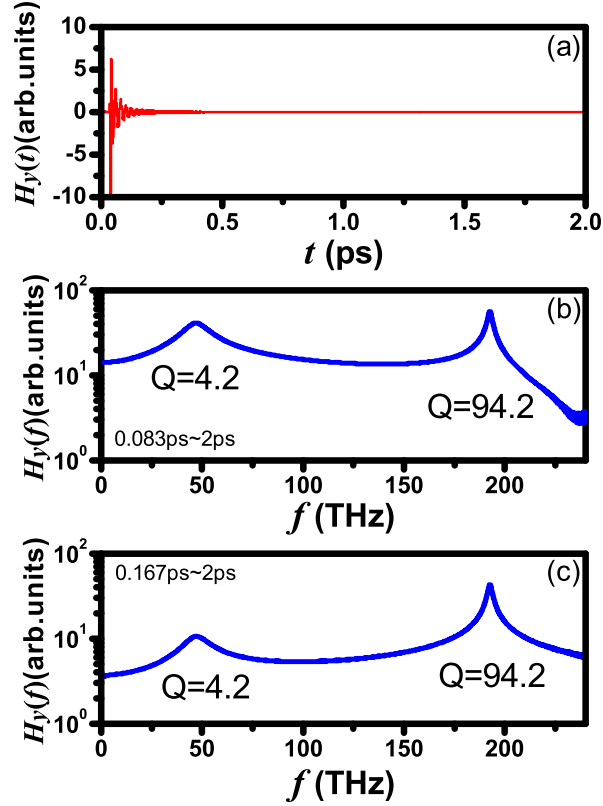}
\end{center}
\caption{(a) The time dependence of the magnetic fields of reflection. Spectra
are acquired through Fourier transformation of the time-domain results with
different time spans (b) $t=0.083\sim2$ps and (c) $t=0.167\sim2$ps.}%
\end{figure}

\section{Rate equation analysis: lasing threshold and dynamic processing}

\begin{figure}[ptb]
\begin{center}
\includegraphics[
width=8cm
]{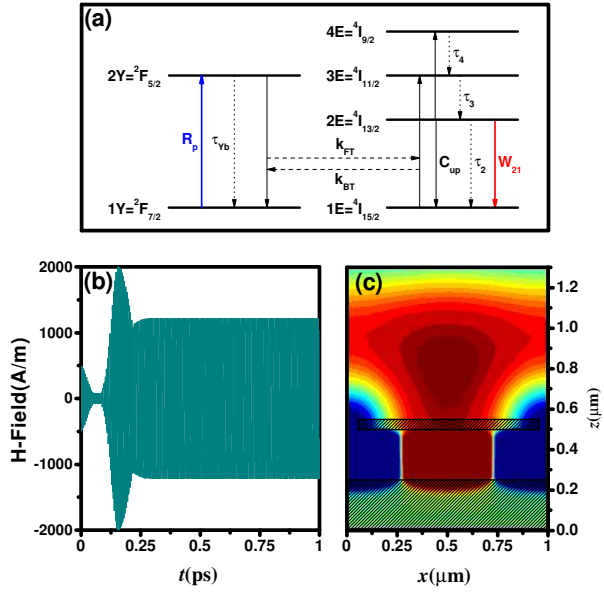}
\end{center}
\caption{(a) Energy level diagram for the Ytterbium-erbium co-doped material
Er:Yb:YCOB. (b) Time dependence of the averaged $H_{y}$ over all the spatial
points of a $xz$ section of the cavity. (c) the magnetic field $H_{y}$
distribution at a typical time ($t=$0.96ps).}%
\end{figure}

Now we investigate the threshold condition of our lasing system by using the
realistic material parameters. A widely used gain medium, Ytterbium-erbium
co-doped material Er:Yb:YCOB is adopted as the active medium for lasing around
$1.5\mu$m\cite{18,19}$.$ Its level assignment is shown in Fig4 (a). The
population density on every level evolves according to the dynamic rate
equations\cite{20}, and can be solved for steady-state conditions:%
\begin{align}
\frac{dN_{\text{2Y}}}{dt}  &  =-\frac{dN_{\text{1Y}}}{dt}\label{eq1}\\
&  =R_{p}-\frac{N_{\text{2Y}}}{\tau_{\text{Yb}}}-k_{\text{FT}}N_{\text{2Y}%
}N_{\text{1E}}+k_{\text{BT}}N_{\text{1Y}}N_{\text{3E}}=0,\nonumber
\end{align}

\begin{gather}
\frac{dN_{\text{1E}}}{dt}=k_{\text{BT}}N_{\text{1Y}}N_{\text{3E}}%
-k_{\text{FT}}N_{\text{2Y}}N_{\text{1E}}+W_{21}N_{l}\Delta N_{21}%
+C_{\text{up}}N_{\text{2E}}^{2}\nonumber\\
+\frac{N_{\text{2E}}}{\tau_{2}}+\beta_{31}\frac{N_{\text{3E}}}{\tau_{3}}%
+\beta_{41}\frac{N_{\text{4E}}}{\tau_{4}}=0, \label{eq2}%
\end{gather}

\begin{align}
\frac{dN_{\text{2E}}}{dt}  &  =\beta_{32}\frac{N_{\text{3E}}}{\tau_{3}}%
+\beta_{42}\frac{N_{\text{4E}}}{\tau_{4}}-\frac{N_{\text{2E}}}{\tau_{2}%
}-W_{21}N_{l}\Delta N_{21}-2C_{\text{up}}N_{\text{2E}}^{2}\nonumber\\
&  =0, \label{eq3}%
\end{align}

\begin{equation}
\frac{dN_{\text{3E}}}{dt}=k_{\text{FT}}N_{\text{2Y}}N_{\text{1E}}%
-k_{\text{BT}}N_{\text{1Y}}N_{\text{3E}}+\beta_{43}\frac{N_{\text{4E}}}%
{\tau_{4}}-\frac{N_{\text{3E}}}{\tau_{3}}=0, \label{eq4}%
\end{equation}

\begin{equation}
\frac{dN_{\text{4E}}}{dt}=C_{\text{up}}N_{\text{2E}}^{2}-\frac{N_{\text{4E}}%
}{\tau_{4}}=0, \label{eq5}%
\end{equation}

$N_{ix}$ and $\tau_{i}$ are the population density and radiative lifetime of
corresponding levels. $\beta_{ij}$ are the branching ratio for transitions
from the $i^{\text{th}}$ to $j^{\text{th}}$ levels. $R_{p}$ is the pump rate
from 1Y to 2Y level. $k_{\text{FT}}$ and $k_{\text{BT}}$ are the forward and
backward energy transfer coefficient corresponding to 2Y$+$1E$\rightarrow
$3E$+$1Y and backwards. $C_{\text{up}}$ is the coefficient of the cooperative
upconversion 2E$+$2E$\rightarrow$4E$+$1E. $W_{21}$ is the stimulated-emission
coefficient. $\Delta N_{21}=N_{\text{2E}}-N_{\text{1E}}$ is the inverted
population difference between 1E and 2E level. $N_{l}$ is the population
density of the lasing mode. Considering the conservation of population
density, we have $N_{\text{Yb}}=N_{\text{1Y}}+N_{\text{2Y}}$ and
$N_{\text{Er}}=N_{\text{1E}}+N_{\text{2E}}+N_{\text{3E}}+N_{\text{4E}}$. The
threshold lasing can be evaluated under steady-state condition as
$\frac{dN_{l}}{dt}=(W_{21}\Delta N_{\text{th}}-\frac{\omega}{Q})N_{l}=0$,
where $Q$ is the Q-factor of the lasing mode. We note that the threshold
inversion density $\Delta N_{\text{th}}$ can be estimated by $Q$; and a higher
Q-factor as a result of lower radiation loss and damping loss of the lasing
mode will give rise to lower threshold doping concentrations or pumping power
dominated by $\Delta N_{\text{th}}$.

The Q-factor of SPP mode can be extracted from the FDTD numerical simulations
by assuming that the dielectric is passive and lossless with a dielectric
constant equal to 1.7, the real part of that of Er:Yb:YCOB. A one-way TM
polarized plane wave with a Gaussian distribution in time domain is normally
incident on the model slab. The incident plane is $2.5\mu$m away from the top
surface of the model slab. A probing plane is positioned $0.35\mu$m behind the
incident plane to monitor the reflected wave fields. Figure 3(a) illustrates
the time-domain spectroscopy of the magnetic local field $H_{y}$ at the
probing plane. The frequency spectra in Figs. 3(b) and 3(c) present the
Fourier transformed data of the time-domain spectroscopy in different time
spans $t=0.083\sim2$ps and $t=0.167\sim2$ps. A broad peak and a narrow one,
centered at 45.0THz and 192.6THz respectively, are observed in both figures.
More calculations in time domain indicate that the decay rate of the broad one
is much quicker than that of the narrow one. These two peaks correspond to the
localized SPP resonance in fundamental mode at 46.3THz and the spatially
coherent SPP resonance in harmonic mode at 193.5THz, as shown in Figs. 1 and
2. Q-factor of a SPP resonance can be extracted from the FWHM in the frequency
spectra. A low Q-factor of only 4.2 and a high Q-factor of 94.2 are derived at
45.0THz and 192.6THz in the frequency spectra. The Q-factor can also be
estimated with the exponential decay rate of the mode in time domain, as
$Q=2\pi f\tau$, where $\tau$ is the lifetime of the mode. In our case, the
life time of each mode is, $\tau=0.0149$ps and $0.0779$ps, respectively,
reproducing exactly the same values of Q-factor at the two resonance
frequencies as the former method. We note that the Q-factor of spatially
coherent SPP mode is of one order larger than that of localized SPP mode. And
a higher Q-factor will be advantageous in the field amplification in gain
medium and the stimulated emission radiations. By applying the condition of
threshold lasing at 192.6THz to the rate equations, a threshold doping
concentration $1.48\times10^{25}$ ions/m$^{3}$ of Er$^{3+}$ is obtained, which
falls inside the typical concentration range of Er$^{3+}$ ions in Er:Yb:YCOB.
In our calculations, the Yb$^{3+}$ concentration is assumed to be fixed at
$5.0\times10^{27}$ ions/m$^{3}$. It is not likely to utilize a fundamental SPP
mode of the meta-surface to realize plasmon lasing due to the low Q-factor of
4.2, as the required Er$^{3+}$ concentration $1.3\times10^{27}$ ions/m$^{3}$
is hard to achieve in Er:Yb:YCOB. More calculations show that it is not very
easy for the Q-factor of the local SPP resonance in fundamental mode to exceed
a value of 10 by tuning structural parameters.

Now we perform FDTD simulations to dynamically demonstrate a lasing example of
our model slab$\cite{21,22}$. Periodic boundary is adopted in our simulations.
the Yb$^{3+}$ concentration and the Er$^{3+}$ concentration are chosen at
5.0$\times$10$^{27}$ irons/m$^{3}$ and 4.0$\times$10$^{25}$ irons/m$^{3}$,
respectively. Calculations are performed by assuming a pumping power at
$P_{\text{pump}}=2.1\times10^{-5}$W which is beyond the threshold
$P_{\text{th}}=1.9\times10^{-5}$W\emph{ }. The magnetic field $H_{y}$ pattern
of the cavity and its emission field distribution at a typical time
($t=0.96$ps) are shown in Fig. 4(c). Fig. 4(b) shows the time dependence of
the averaged $H_{y}$ over all the spatial points in $xz$ plane covering the
unit cell.

\begin{figure}[ptb]
\begin{center}
\includegraphics[
width=8cm
]{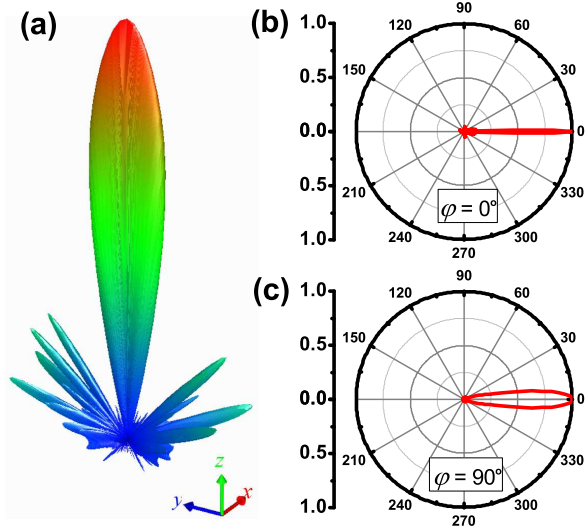}
\end{center}
\caption{Far-field radiation patterns $\left\vert E\right\vert ^{2}$ at
192.6THz as a function of polar angle $\theta$ and azimuthal angle $\varphi$.
(a) $\left\vert E(\theta,\varphi)\right\vert ^{2}$ in a stereogram chart, (b)
$\left\vert E(\theta,0^{\text{o}})\right\vert ^{2}$ (E-plane), (c) $\left\vert
E(\theta,90^{\text{o}})\right\vert ^{2}$\ (H-plane).}%
\end{figure}

The lasing at 192.6THz of a real structure is also considered. In the FDTD
simulations, the lateral size of the model slab is 20 periods along $x$
direction (the direction of the grating) $w_{x}=20p$, and the length of the
stripe along $y$ direction is $w_{y}=w_{x}$. The angular dependence of
far-field patterns are calculated at the steady-state. We see from Fig. 5 that
most of the energy of the lasing beam is confined in a small solid angle. The
angle-dependent power is normalized to the maximum at the normal direction.
The FWHM in the E-plane ($\varphi=0^{\text{o}}$) is $\Delta\theta=4^{\text{o}%
}$, and that in the H-plane ($\varphi=90^{\text{o}}$) is $\Delta
\theta=16^{\text{o}}$. The highly directional lasing beam shall be attributed
to the excitation of a spatially coherent SPP mode on the slab, as such an SPP
mode will inherently establish phase correlations among the unit cells on the
meta-surface due to the dominant $\pm$1$^{\text{st}}$ orders of guided Bloch
modes in the active layer.

\section{Lasing in the visible region}

Our lasing system can also operate at visible wavelength with an appropriate
design. Figure 6 presents an example of reflectance amplification at 474 THz
($\lambda=633$nm). A rather low gain coefficient of $\alpha_{\text{th}}%
=267$cm$^{-1}$ is enough for the compensation of metal dissipation. Even the
optimal gain coefficient, $\alpha_{\text{opt}}=857$cm$^{-1}$ for the maximum
reflectance, falls in the range of the gain coefficients of the most commonly
used active media in this frequency regime, such as organic dye molecules,
semiconductor quantum wells, and nano-crystal quantum dots. The structural
parameters are as follows: grating period $p=400$nm, metallic strip width
$a=385$nm, air gap width $g=15$nm, grating thickness $t=20$nm, and dielectric
layer thickness $h=100$nm. This calculation demonstrates that the spatial
coherence of a harmonic magnetic SPP resonance can be harvested for plasmon
lasing in the visible region.

\begin{figure}[ptb]
\begin{center}
\includegraphics[
width=8cm
]{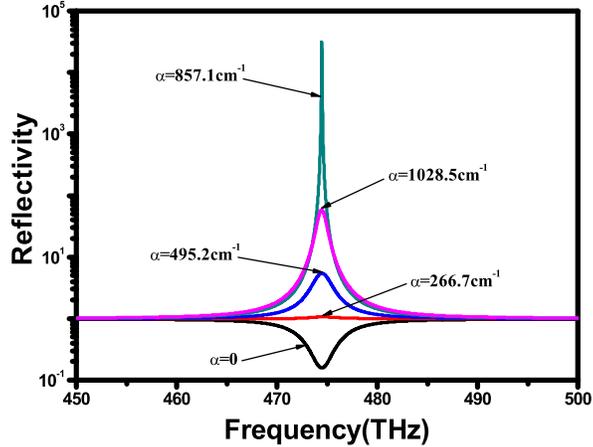}
\end{center}
\caption{The spectra of reflection amplification around 474THz with respect to
different values of gain coefficient $\alpha$\ of the active medium.}%
\end{figure}

\section{Conclusion}

In summary, we demonstrate that spatially coherent SPP modes in harmonic mode
on a planar plasmonic meta-surface can be utilized for low-threshold
directional plasmon lasing. As both the Q-factor of the spatially coherent SPP
modes and the confinement of the local field are dictated primarily by
structural parameters, the SPP amplification can be triggered at a rather low
threshold gain strength attainable with most of the active media. Meanwhile
our examples indicate that the planar low-threshold plasmon lasing is not
difficult to be realized with current fabrication techniques, even in the
visible region. Our calculations show that using the harmonic SPP modes is
crux of the matter for achieving low-threshold lasing on a magnetic
meta-surface rather than using the fundamental ones.

\begin{acknowledgments}
This work was supported by NSFC (No. 10974144, 60674778), CNKBRSF (Grant No.
2011CB922001), HK RGC grant 600308, the National 863 Program of China (No.
2006AA03Z407), NCET (07-0621),STCSM, and SHEDF (No. 06SG24). XY Jiang and W Li
acknowledge NSFC (Grant No. 11004212, 10704080, 60877067 and 60938004) STCSM
(Grant No. 08dj1400303) and SNSFC(Grant No. 11ZR1443800).
\end{acknowledgments}

\end{document}